# Domain Wall-Magnetic Tunnel Junction Analog Content Addressable Memory Using Current and Projected Data

Harrison Jin, Hanqing Zhu, Keren Zhu, Thomas Leonard, Jaesuk Kwon, Mahshid Alamdar, Kwangseok Kim, Jungsik Park, Naoki Hase, David Z. Pan, and Jean Anne C. Incorvia

*Abstract*—With the rise in in-memory computing architectures to reduce the compute-memory bottleneck, a new bottleneck is present between analog and digital conversion. Analog content-addressable memories (ACAM) are being recently studied for in-memory computing to efficiently convert between analog and digital signals. Magnetic memory elements such as magnetic tunnel junctions (MTJs) could be useful for ACAM due to their low read/write energy and high endurance, but MTJs are usually restricted to digital values. The spin orbit torque-driven domain wall-magnetic tunnel junction (DW-MTJ) has been recently shown to have multi-bit function. Here, an ACAM circuit is studied that uses two domain wall-magnetic tunnel junctions (DW-MTJs) as the analog storage elements. Prototype DW-MTJ data is input into the magnetic ACAM (MACAM) circuit simulation, showing ternary CAM function. Device-circuit co-design is carried out, showing that 8-10 weight bits are achievable, and that designing asymmetrical spacing of the available DW positions in the device leads to evenly spaced ACAM search bounds. Analyzing available spin orbit torque materials shows platinum provides the largest MACAM search bound while still allowing spin orbit torque domain wall motion, and that the circuit is optimized with minimized MTJ resistance, minimized spin orbit torque material resistance, and maximized tunnel magnetoresistance. These results show the feasibility of using DW-MTJs for MACAM and provide design parameters.

## I. INTRODUCTION

AS data size increases and Moore's law slows down, alternative in-memory computing (IMC) architectures are being studied and used to efficiently process information. Conventional analog IMC often uses crossbar array architectures and nonvolatile memory (NVM) such as resistive random-access memory (RRAM) and has shown many applications in neural network acceleration [1]–[3], neuromorphic computing [4], statistical learning [5], signal processing [6], [7], scientific computing [8], [9], and other fields. However, there is a challenge extending these architectures to operations other than matrix dot multiplication. Existing IMC architectures rely on extensive analog to digital conversion (ADC) and digital to analog conversion (DAC) to switch between matrix multiplication and other computations such as activation functions [10]. The consequent conversion cost greatly dilutes efficiency in both power and speed; e.g., data converters can consume around 85% of the total energy in a typical RRAM-based neural network accelerator [11]. This energy overhead is especially critical for edge computing.

To address this challenge, analog content-addressable memories (ACAM) are being recently studied for IMC [12], [13]. ACAM cells are designed to detect whether the value corresponding to input search data is located within a range. Unlike conventional CAM, the input data of ACAM are allowed to be analog values or multi-bit. ACAM can be used to fuse computation and data conversion for time- and energy-efficient conversion between analog and digital signals. For example, a recently proposed RRAM-based ACAM shows delays of 350 ps and >1 fJ energy consumption, and a recently shown ferroelectric-based ACAM shows up to 3-bit precision to perform in-memory nearest neighbor searching to perform few-shot learning [14], [15]. The ideal requirements of NVM for ACAM include low switching energy, low read energy, high endurance, and controllability of setting the memory element to a given analog value. Thus, magnetic tunnel junctions (MTJs) are a natural choice for ACAM, due to their theoretically unlimited endurance, modest switching voltage, and back-end-of-the-line compatibility for integration into the ACAM cell [16]. The domain wall-magnetic tunnel junction (DW-MTJ) allows for analog-like programming of tunnel magnetoresistance (TMR) through modulation of the position of a DW underneath an MTJ, using either spin transfer torque (STT) or spin orbit torque (SOT). Recent work has demonstrated the use of STT-based MTJs in ternary CAM applications, but ternary CAMs are associated with considerably greater area consumption costs in order to achieve the same density of bits as their contemporary analog and multi-bit counterparts [15], [17], [18]. STT magnetic random-access

This work was supported by the Samsung Global Research Outreach (GRO) Program. The authors also acknowledge support from the Department of Energy Office of Science Microelectronics Co-Design project COINFLIPS (J. Kwon) and the National Science Foundation Graduate Research Fellowship under Grant No. 2020307514 (T. Leonard). *(Corresponding author: Jean Anne C. Incorvia)*.

H. Jin, H. Zhu, K. Zhu, T. Leonard, J. Kwon, M. Alamdar, D. Z. Pan, and J. A. C. Incorvia are with the Department of Electrical and Computer Engineering, University of Texas at Austin, Austin, TX 78712 USA (e-mail: incorvia@austin.utexas.edu). M. Alamdar is now with Samsung Austin Semiconductor, Austin, TX, 78754 USA.

K. Kim, J. Park, and N. Hase are with Samsung Advanced Institute of Technology (SAIT), Samsung Electronics Co., Suwon, 16678, South Korea (e-mail: stone99.kim@samsung.com).

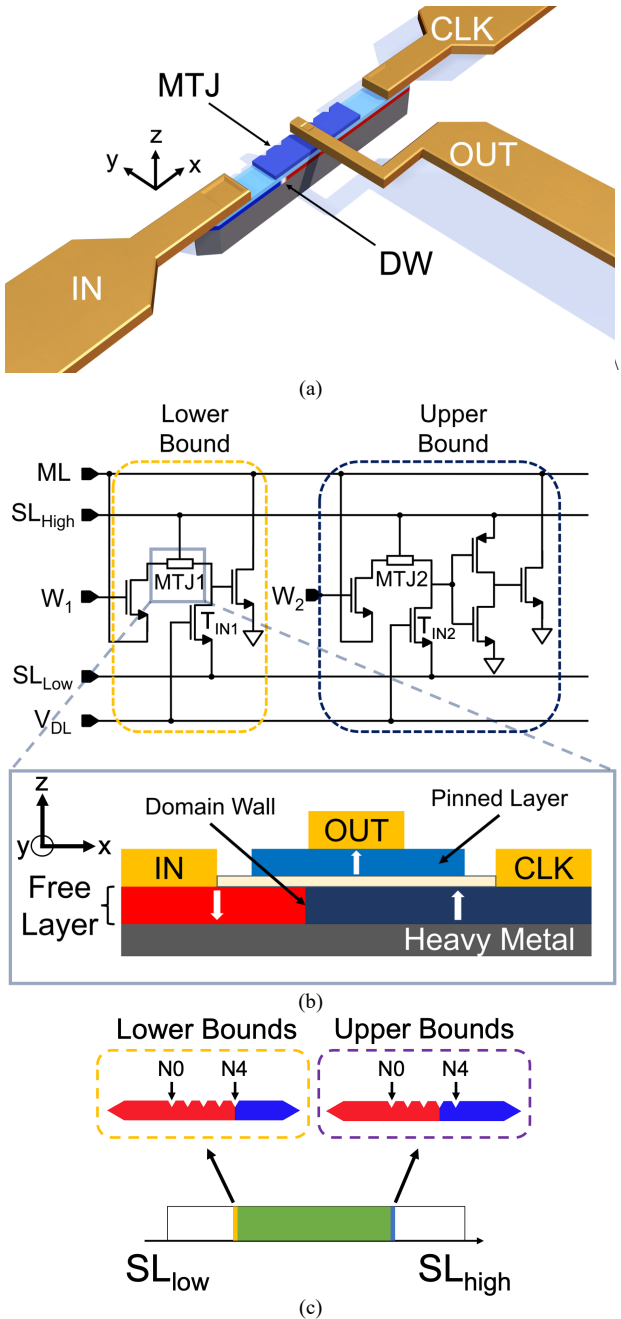

Fig. 1. (a) Diagram of the three-terminal DW-MTJ. Resistance states are programmed using voltage pulses from IN and CLK, and read from IN to OUT. Notches are shown that assist repeatable setting of the DW. (b) Circuit schematic of proposed MACAM circuit. Minimum and maximum voltage bounds are set using $SL$ lines, and input search voltage is applied through the data line, $V_{DL}$. The search result is reflected on the match line ($ML$) behavior. (c) Programmable resistance states demonstrate how different voltage states on both MTJs can be used to write different voltage bounds. A match can be yielded anywhere between the upper and lower bounds. Different notch locations are shown on a 5-weight DW-MTJ synapse. "N4" corresponds to antiparallel and "N0" corresponds to parallel resistance relative to the reference

memory (STT-MRAM) based on the MTJ has also been shown in crossbar arrays to perform analog multiply-and-accumulate operations [19]. But, much of the previous works of spintronics for CAM has been focused on binary and ternary functionality. This is because achieving controllable multiple resistance levels in an MTJ is a challenge.

Here, we propose and verify the performance of a DW-MTJ ACAM prototype for high-throughput, high-speed searches. The DW-MTJ ACAM cell compares an analog input to a range of stored values, which is set using the programmable resistance of the DW-MTJ through the modulation of the DW. Circuit simulations using prototype results from DW-MTJ device cycling data verify that MTJs can be effectively integrated into ACAMs as programmable elements. Additionally, projected data using optimized magnetic stack parameters demonstrate the feasibility of performing analog multi-bit search operation for implementation in high-throughput computing. Our proposed DW-MTJ ACAM benefits from up to a 44 × decrease in energy consumption per search, and 2.86 × faster search time, per bit compared to existing MTJ ternary CAM circuits. Additionally, the high programmability in linearity through different DW-MTJ geometries, combined with low variation within each weight, allows our proposed ACAM to circumvent time-costly weight programming that is necessary in other emergent ACAM designs; thus, potentially reducing write times by up to 3 orders of magnitude. These results show the potential for DWs and MTJs to be used in these energy-efficient circuits.

## II. CELL DESIGN AND METHODS

Fig. 1a shows the DW-MTJ multi-weight NVM used for the magnetic ACAM (MACAM) design. A top-pinned MTJ stack has its bottom heavy metal and magnetic free layer extended into a magnetic wire that hosts a magnetic DW. SOT current applied from IN to CLK sets the DW position at one of the notches, which in turn sets the resistance between the CLK and OUT terminals. We have previously shown DW-MTJ prototypes with 3-5 stable resistance levels at room temperature; due to the physical setting of the resistance by the DW position, highly controllable weights are achievable as long as the DW is set to the desired notch [20].

Two DW-MTJs are integrated into the 8-transistor MACAM cell shown in Fig. 1b [18]. Minimum and maximum voltage bounds are set using the search lines ($SL_{High}$, $SL_{Low}$), and input search voltage is applied through the data line, $V_{DL}$. The search result is reflected on the match line ($ML$) behavior. Fig. 1c depicts how the DW position in the DW-MTJ determines a match. Programmable resistance states demonstrate how different voltage states on both MTJs can be used to write different voltage bounds. A match can be yielded anywhere between the upper and lower bounds.

To understand the cause, the drain-to-source voltage of the input transistor $T_{IN1}$ (see Fig. 1b), $V_{DS}$, and its drain current, $I_D$, can be described using the form:

$$I_D = \frac{SL_{high} - V_{DS}}{(R_{wire} + R_{MTJ})}, \quad (1)$$

where $R_{wire}$ is the resistance of the heavy metal plus free layer patterned wire and $R_{MTJ}$ is the read-out resistance of the DW-MTJ. Subsequently,

$$V_{DS} = SL_{high} - I_D[R_{wire} + R_{MTJ}]. \quad (2)$$

The saturation region of the $T_{IN1}$ in which lower bounds will remain matched can then be defined as:

$$V_{DL} > \sqrt{\frac{SL_{high}-V_{Th,PD}}{[R_{wire}+R_{MTJ}]k_n}} + V_{TH,IN}, \quad (3)$$

Where $V_{TH,PD}$ and $V_{TH,IN}$ (in Figs. 3b, e) are the threshold voltages of the pulldown ($T_{PD}$) and input ($T_{IN}$) transistors, respectively. $k_n = \mu_n C_{ox} \frac{W}{L}$, the large signal MOSFET transconductance parameter, where $\mu_n$ is the mobility of electrons on the channel surface, $C_{ox}$ is the oxide capacitance, and $\frac{W}{L}$ is the ratio of channel width to channel length. Meanwhile, the linear region where the lower bounds will remain matched can be defined using the approximation:

$$V_{DL} < \frac{V_{Th,PD}}{g_m}\left[\frac{1}{r_o} + \frac{1}{R_{wire}+R_{MTJ}}\right]. \quad (4)$$

Here, $g_m$ is the small signal transconductance and $r_o$ is the output resistance, both of which are intrinsic constants to the n-type MOSFET, $T_{IN1}$. The opposite can be said about the upper bound, as the gate voltage of the pulldown transistor is inverted.

As the input search voltage decreases to 0, so does the drain current, $I_D$, of the input transistor, as is expected by the $V_{GS}$-$I_D$ relationship of a n-type MOSFET. In the ACAM circuit, $V_{GS}$, the gate-to-source voltage of the input transistor, is equal to the analog search input, $V_{DL}$. To maintain the matching condition, it is crucial that $V_{DS}$ remains less than that of $V_{Th,PD}$ in order to maintain the match line voltage. To counteract this, the MTJ resistance must increase accordingly to maintain the adequate voltage drop across the MTJ to keep $T_{PD}$ in its OFF state.

Cadence Virtuoso and Spectre are used to characterize the functionality of the MACAM circuit. The circuit is constructed using 40 nm gate processes technology, and the MTJ is modeled as two resistances in series, $R_{wire} + R_{MTJ}$. To verify performance, a two-dimensional parametric sweep of search voltages at different $R_{MTJ}$ is run. The input search voltage is swept from the full range established by the search lines ($SL_{Low}$ and $SL_{High}$ set to 0 V and 1 V respectively) at a preprogrammed $R_{MTJ}$. This is repeated at different $R_{MTJ}$ to determine the ML behavior as a function of the search inputs, to understand the limits of both the DW-MTJ device and CMOS circuitry.

III. TCAM FUNCTIONALITY WITH PROTOTYPE DATA

To start, experimental data from DW-MTJ prototypes is input into the constructed circuit model, to study their function for CAM. Fig. 2a shows the device data with device SEM shown in the inset. A 50 ns voltage pulse is applied from IN to CLK, followed by measurement of $R_{MTJ}$. The voltage pulse amplitude is increased from 2 to 4 V in 0.1 V steps, showing three distinct resistance values as the DW eventually de-pins and moves to another notch. This is repeated for 10 cycles; see Ref [20] for details. Nominal TMR of the magnetic stack was measured using current in-plane TMR = 170%. The resistance-area product for parallel MTJ resistance, RA, was measure RA = $\Omega \times \mu m^2$, with a heavy metal layer of tantalum. The trapezoidal synapse device used an MTJ with top-down area of 1.575 $\mu m^2$.

From this data, we extract total resistances of $R_{MTJ}$ = 67 Ω, 75 Ω, and 93 Ω, with cycle-to-cycle resistance variation = 2.5%.

Inputting these values into the MACAM circuit, ternary CAM behavior is seen, shown in Fig 2b, which plots the ML voltage vs. search voltage for the different relative MTJ weights. The search voltage is the analog search input from $V_{DL}$. From these curves, the lower bound $B_L$ (V) and upper bound $B_U$ (V) are defined as the search voltage values that set $ML = 0.5$. The storage range is defined as $SR = B_U - B_L$. The resulting maximum $SR = 0.109$ V that can be achieved using these measured resistance weights is between $B_L = 0.854$ V and $B_u = 0.963$ V, which can be seen as the *don't care* or *X*, state that includes all values within this range; alternatively, by setting the upper and lower bounds DW-MTJs to the same weight, we can also achieve a cell which will always result in a mismatch. Because there are 3 resistance weights, it is also possible to achieve two smaller resistance states as well, which can be used in binary implementation, depicted as the 0 and 1 states in Fig. 2b. The combination of the *X* bit with the smaller 0 and 1 bits can then be used to implement ternary CAM functions. Ternary CAM application in memory-augmented neural networks have previously been demonstrated for one-shot learning in Ref. [21].

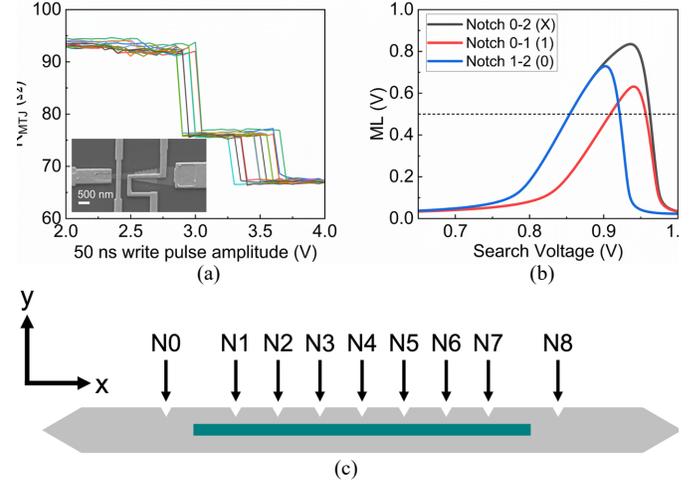

**Fig. 2.** (a) Data of $R_{MTJ}$ vs. applied voltage pulse amplitude of device shown in inset, showing 3 distinct weights. Each color is another cycle of the same device showing repeatable cycle-to-cycle behavior. (b) Calculated ternary CAM performance of device from (a). The dotted line at 0.5 V on the ML shows the minimum ML voltage necessary to yield a match. The lower and upper bound of each discrete level (0, 1, or *X*) is marked by the two points of intersection with the dotted line. (c) Top-down design of 9 weight DW-MTJ wire, with lithographically patterned magnetic wire shown in grey with 9 notches, and the MTJ for read-out is shown in teal.

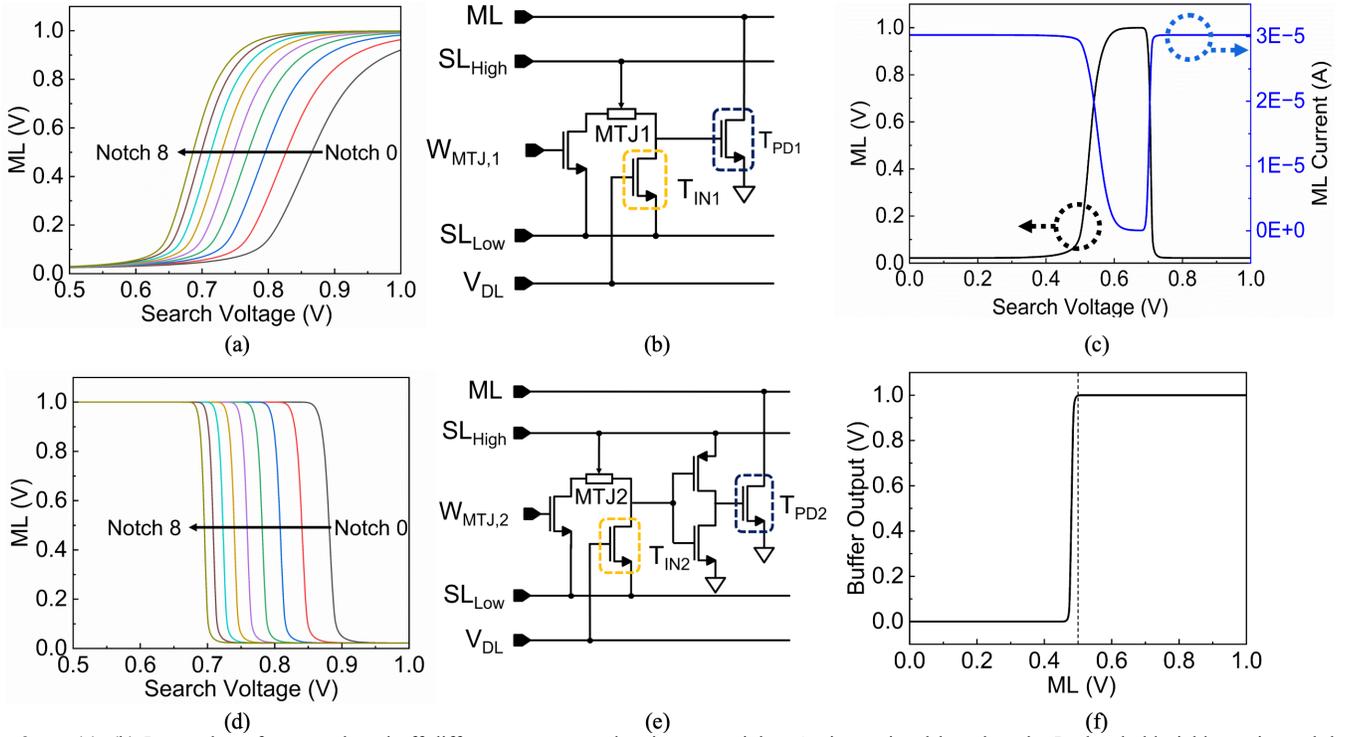

**Fig. 3.** (a), (b) $B_L$ match performance based off different programmed resistance weights. An input signal less than the $B_L$ threshold yields a mismatch by forcing the transistor PD₁ in (b) to pull down the ML voltage to ground. (c), (d) Conversely, an input signal greater than the $B_U$ threshold in (d) forces the transistor PD₂ to pull down ML voltage to ground using similarly to (c) but with an inverting CMOS prior to the pull-down transistor PD₂. (e) Simulation of voltage and current measured through the match line during an input voltage sweep. (f) Transfer characteristics of voltage buffer at the match line output indicating a matching threshold at 500 mV.

## IV. ACAM FUNCTION AND DESIGN FOR 9-RESISTANCE WEIGHT DW-MTJ

With existing prototypes showing 3-5 resistance levels, it is feasible to extend to 8 resistance levels by extending the length of the DW track and including 9 notches, depicted in Fig. 2c. Assuming $R_{wire} = 40\ \Omega$, and $R_{MTJ} = 70\ \Omega$, with TMR = 170%, more analog and multi-bit capabilities of the cell can be demonstrated.

Under these assumptions, we can simulate the performance of $B_L$ and $B_U$'s circuit components. The specific voltage value of the bound associated to the $B_L$ circuit can be programmed using the MTJ-transistor voltage divider circuit, which is demonstrated in Figs. 3a, b. Furthermore, it can be seen that the relationship between the DW-MTJ resistance weights and their associated bounds is such that lower resistances are necessary to write higher bounds, while larger resistances are required to achieve the lower bounds, shown in Figs. 3a, d where the parallel resistance is in notch "0" and increases with each notch up to notch "8", which is the anti-parallel state. $B_U$ (see Figs. 3d, e) experiences a similar matching condition, but conversely requires an input voltage smaller than the threshold at $T_{PD2}$ due to the CMOS inverter prior to the pulldown transistor. Fig. 3c depicts the match line current (blue) and match line voltage (black) during an input voltage sweep from 0 to 1 V, depicting a mismatch-match-mismatch event. To assess the *ML* threshold voltage to yield a match, a buffer is placed at the end of the *ML*, and its transfer function is shown in Fig. 3f, revealing a minimum *ML* voltage at 0.5 V to be considered a match.

When the *ML* is plotted against the search voltage for the 9 evenly spaced notches in the DW-MTJ, the resulting relationship is 9 unevenly spaced discrete levels, as shown in Fig. 4a. While ACAM behavior is still achieved, it is shown that linearly spaced notches produce uneven widths of the distinct levels in the multi-bit circuit operation. With these matching characteristics, it would not be suitable to implement multi-bit performance, such as that shown in Ref. [15]. Thus, designing the DW-MTJ with unevenly spaced notches, to achieve approximately exponentially increasing MTJ resistance vs. notch, alleviates this issue. When considering notch spacing, the minimum pitch between notches is no less than the width of the notch itself to avoid stochastic movement of the DW between adjacent notches from factors like variations in magnetic wire geometry and thermal effects (i.e., a notch with a 100 nm width is restricted to have a minimum spacing of 100 nm from the next notch). Fig. 4b shows the re-designed 9 notches, and the resulting ACAM function of ML vs. search voltage, which shows evenly spaced discrete levels. These results show a useful benefit of DW-MTJs for these circuits because device behavior can be tuned to the circuit by adjusting device geometry.

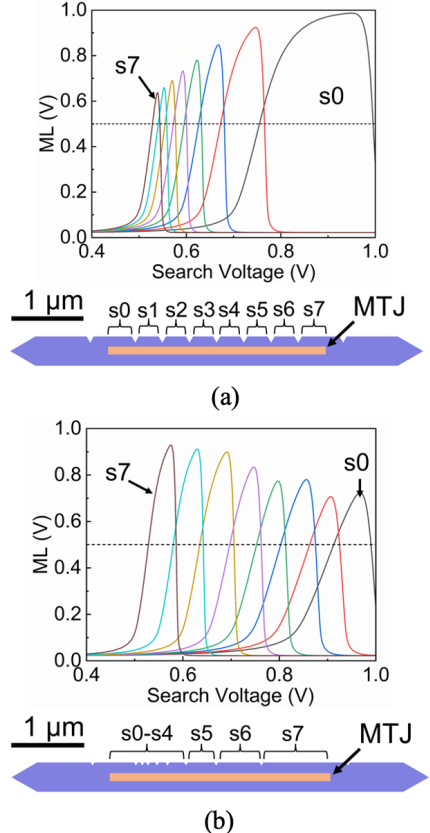

**Fig. 4.** (a) Linear notch spacing vs. (b) nonlinear notch spacing effects on 9-weight multi-bit performance.

## V. MACAM Using MTJ Wafer Data

While the previous section focused on the impact of $R_{MTJ}$ on the ACAM function, the resistance of the heavy metal plus magnetic track, $R_{wire}$, will also impact the circuit performance, since the read current of the DW-MTJ runs through the DW racetrack wire and out the MTJ. For SOT-driven DW motion, $R_{wire}$ is dominated by the resistivity of the heavy metal underneath the DW track. Here, we consider 3 common heavy metals used in SOT-MRAM: platinum, $\alpha$-tungsten, and tantalum; $\beta$-tungsten and other similar large spin Hall angle materials were not included due to their known high resistivities [22]–[24].

### A. Magnetic Stack Material and Device Characteristics

SOT-MRAM thin film stacks were grown with heavy metal layer of 7 nm-thick α-tungsten and measured using CIPT, showing average TMR = 170%, and average RA product = 35 $\Omega \times \mu m^2$. Using these measured stack characteristics, we then evaluated the impact of heavy metal resistivity on device performance. $R_{wire}$ was calculated using the excess length of wire outside the area of the MTJ, with $l \times w$ dimensions of $0.75\ \mu m\ \times 400\ nm$ on all 3 stacks. The resistivities of the heavy metal thin films used were assumed from literature to be 15 $\mu\Omega \times cm$ for platinum, 21 $\mu\Omega \times cm$ for $\alpha$-tungsten, and 25 $\mu\Omega \times cm$ for tantalum [25]–[27]; thus, resulting in a projected wire resistance of 40 Ω, 56 Ω, and 67 Ω, respectively. The top-down geometry of the MTJ has a $l \times w$ dimensions of 3 $\mu m\ \times 100\ nm$, resulting in a parallel resistance of ~117 Ω.

### B. Simulated Results

Fig. 5 shows the performance of the circuit simulated using device parameters extrapolated from each of the 3 stacks. Fig. 5a shows the maximum storage range of all 3 of these devices plotted against each other. The storage range, $SR$, is the maximum distance that can be achieved between $B_U$ and $B_L$. Achieving larger $SR$ allows for greater density of discrete levels in analog multi-bit applications. Additionally, maximizing the accessible storage range within the minimum and maximum possible bounds, established by $SL_{high}$ and $SL_{low}$, reduces the energy cost of peripheral circuitry used to scale down that of the two DW-MTJ-CMOS subcircuits. This is important because the limited TMR ratios available in current MTJ devices allows programming bounds to only a fraction of the total available range. For the three SOT material types, Fig. 6a shows the MACAM bound (V) associated with different values of $R_{MTJ}$. The presence of wire resistance results in unwanted static voltage drops within the subcircuits shown in Figs 3b, e. Due to the low resistivity of platinum thin films, platinum has the highest $B_U$ overall due to having the lowest wire resistance, seeing as both $B_L$ and $B_U$ increase with decreasing MTJ resistance. Consequently, the maximum $SR$ of the ACAM cell utilizing the platinum stack is 245 mV, which is 16% greater than α-tungsten and 28% greater than tantalum. Another demonstration showing the ability to achieve 5 discrete levels can be seen in Figs. 5b-d. Within the range of voltages available to all three stacks, they are all capable of comfortably fitting 5 discrete levels; that is, the minimum pitch between notches is reliably spaced as described in Section IV.

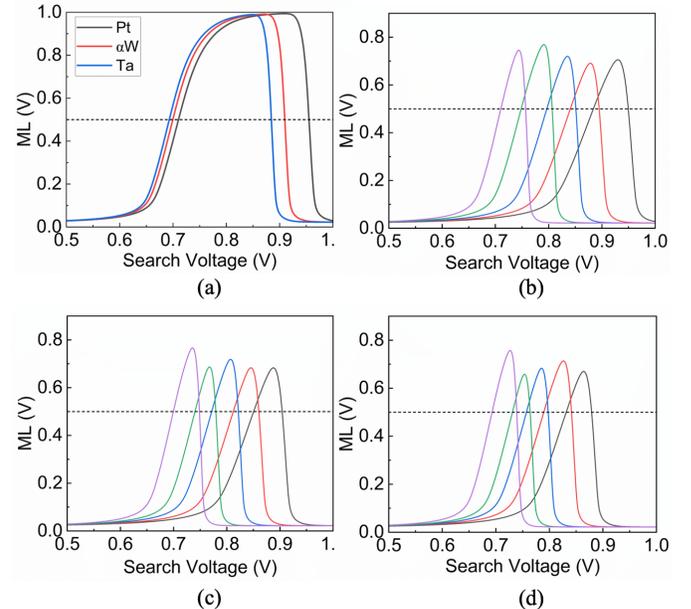

**Fig. 5.** (a) Simulation of three different SOT heavy metals showing total range, as well as individual states assuming 5 notches in the (b) platinum stack, (c) $\alpha$-tungsten stack, and (d) tantalum stack.

## VI. DW-MTJ Materials Parameters Optimization for ACAM

The results so far show that the MACAM circuit can achieve both ternary and multi-bit-like functionality, using prototype data, measured MTJ stack data with often-used heavy metal materials, and feasible extension from the measured 3-5 notches to 9 notches. Here, we inspect design considerations of the DW-MTJ to further optimize the ACAM cell's performance, to predict what the ideal properties of the DW-MTJ should be for this application.

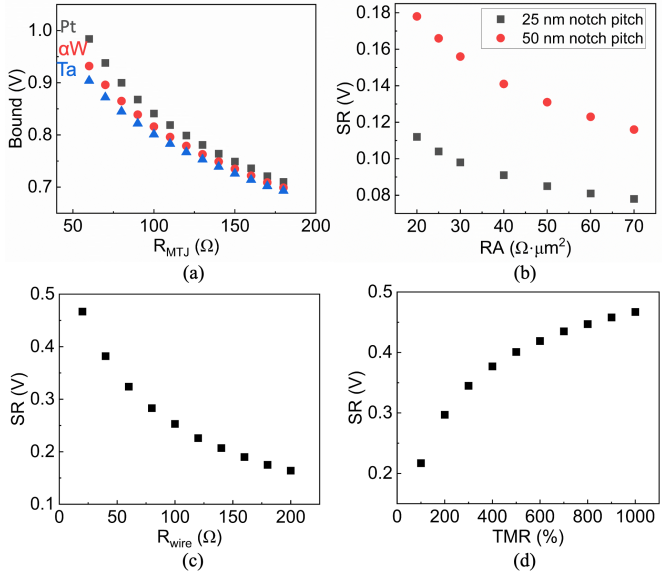

**Fig. 6.** (a) Relation of ML threshold bound vs. $R_{MTJ}$ for the three heavy metal types. (b) Performance changes from proportional geometric scaling of device with 50 nm and 25 nm pitch between notches. (c) Change in maximum writable *SR* of platinum-based MACAM device with wire resistance. (d) Change in maximum writable storage range with TMR.

### A. Stack Characteristics and DW-MTJ Geometry Considerations in Design Optimization

When considering the factors important to optimizing the DW-MTJ for greater storage range in the ACAM cell, stack characteristics (RA, TMR, and resistivity) and device geometry (heavy metal thickness and MTJ top-down area) play large roles in minimizing resistance losses and increasing total storage range. Figs. 6b, c show the storage range decreases with increasing RA and increasing $R_{wire}$; Fig 6d shows storage range improves with TMR with decreasing benefits for high TMRs. The area of the MTJ with respect to stack RA cannot be neglected, since proportionally scaling down a device can have unintended consequences on the total storage range of said device. Fig. 6b shows proportionally scaling down a device with feature node size of 50 nm down to 25 nm results in a subsequent $4\times$ increase of parallel resistance, as both the length and width of the MTJ are each proportionally scaled down by $\frac{1}{2}\times$. The resistance vs. feature node can be accounted for in the circuit design.

### B. Design and Simulation of Prototype with Projected Data

Thus, we design a theoretical MTJ stack with ideal parameters to demonstrate projected prototype performance. Platinum is chosen as the heavy metal due to its low resistivity while still having a good spin Hall angle for energy efficient DW motion [21]. The Pt thin film layer is assumed to be 15 nm thick, and the RA is assumed to be 5 $\Omega \times \mu m^2$, on the low end of what is feasible with today's MgO-based MTJs. We first consider a reasonable TMR = 200%, which is currently achievable. The device is designed to accommodate an MTJ with dimensions of 1.5 $\mu m \times 50\ nm$ to be able to make use of 25 nm pitch between notches. With this, the parallel resistance of the device is 67 $\Omega$ and the anti-parallel resistance is 200 $\Omega$. Fig. 7 shows the simulation results, where the ACAM is demonstrating 8 and 10 discrete levels by choosing 9 or 11 notches respectively, or also a minimum resolution of 3-bits. The trends observed in Fig. 7 reveal that large TMR and low RA work to improve the maximum storage range of the ACAM cell: the storage range increases from 245 mV, projected from the realistic magnetic stacks in the previous section, up to 300 mV in the ideal stack. The increase in heavy metal layer thickness from 7 nm to 15 nm constitutes a decrease in wire resistance from 40 $\Omega$ to 19 $\Omega$. This, in combination with the 30% increase of TMR, extends the projected storage range by ~18%. Given the nonlinear behavior of wire resistance, shown in Figs. 4 & 5, the increased range of programmable resistances, and their associated voltage bounds, allow for larger density of notch spacing necessary in the lower discrete levels for multi-bit implementations. Fig. 7 shows 8 and 10 discrete levels, with devices designed such that they meet the design considerations for minimum notch spacing described in Section IV.

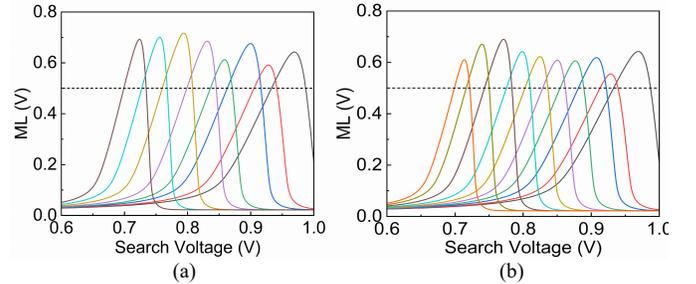

**Fig. 7** (a) DW-MTJ utilizing platinum heavy metal layer with ideally optimized RA, scaled geometry, 200% TMR, and minimal wire resistance to demonstrate multi-bit performance of 9 notches and 8 distinct levels. (b) Identical parameters assumed for 11 notches and 10 distinct levels.

### C. Simulation of Prototype with 1000% TMR Ratio Projected Data

In Fig. 8, the same parameters are assumed except TMR = 1000%, not yet commercially achievable today. The increase in storage ranges from 200% TMR to 1000% TMR is 300 mV to 480 mV. If this high on/off ratio could be achieved, the cell would be capable of greater multi-bit precision, as much as 16 discrete levels, or 4-bits, as shown in Fig. 7d.

## VII. ANALYSIS AND DISCUSSION

To evaluate the energy consumption of the MACAM, we simulate the average DC current through the *ML* and integrated it over several inference passes. Using this method, the estimated energy consumption during one search period in our cell is roughly 0.92 fJ per search operation. It should be noted, however, that the energy consumption from periphery circuits (match line pre-charging, search line drivers, DAC, etc.) is estimated to consume up to an additional 0.52 fJ [18]. To estimate the total area consumption of the CMOS components, the total sum of all transistors was taken and assumed to be ~90% of the total area consumption; thus, giving a top-down circuit area consumption of ~36 $\mu m^2$ using a 40 nm CMOS technology node. The largest dimension of MTJ devices used in previous simulations does not exceed 5.54 $\mu m^2$; thus, DW-MTJ placement back-end-of-the-line on the CMOS circuit would not affect the overall top-down area of the circuit.

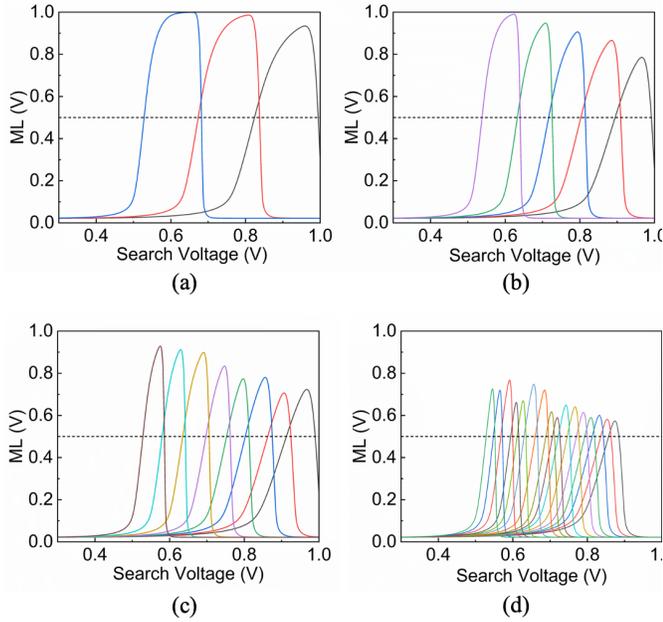

**Fig. 8.** (a) DW-MTJ utilizing platinum heavy metal layer with ideally optimized RA, scaled geometry, 1000% TMR, and minimized wire resistance to demonstrate multi-bit performance with 3 weights. (b-d) Identical parameters are used to extend the resolution up to 5, 8, and 16 states.

TABLE I
COMPARISON OF EMERGENT ACAM ARCHITECTURES*

| PARAMETER | FEFET [28] | RRAM [18][29] | SRAM [18] | DW-MTJ [30] |
|---|---|---|---|---|
| AREA CONSUMPTION | 49 $F^2$ | 48,828 $F^2$ | 918,750 $F^2$ | 22,875 $F^2$ |
| TECHNOLOGY NODE | 45 nm | 16 nm | 16 nm | 40 nm |
| NON-VOLATILITY | Yes | Yes | No | Yes |
| ON/OFF RATIO | $10^4$-$10^6$ | $10^6$ | $10^6$ | 1.5-6 |
| VARIATION | High | High | Low | Low |
| LINEARITY | Low | Low | High | High |
| SEARCH LATENCY | ~10 ns | ~50 ps | N/A | 350 ps |
| ENDURANCE | $10^5$ | $10^{12}$ | >$10^{15}$ | >$10^{15}$ |
| ENERGY (PER SEARCH) | 0.07 fJ | 0.52 fJ | 0.165 fJ | 0.92 fJ |

*Values from individual cell

Some additional energy costs can be found in the necessary circuits to perform read and write operations within the ACAM cell. The energy required to update the DW-MTJ by a single weight is on average ~0.1 pJ in few-100 nm prototypes [20], which can be scaled to ~2 fJ for 15 nm feature sizes [31]. This energy can be reduced through scaling and device engineering. The match line output also requires a sensing circuit based on a transimpedance amplifier (TIA), which has an associated energy dissipation of about 2.5 pJ over the course of one search operation [32]. This relatively high energy can be effectively reduced using large ACAM arrays to amplify integrated currents.

The modest TMR of MTJs are considerably smaller than that of the relatively large on/off conductance ratios of FeFETs and RRAM. The operation in the relatively small range of conductivities leads to reduced noise robustness and potential energy costs to scale voltage inputs to a range that can be accommodated by the MACAM. However, FeFETs and modern memristor technology continue to suffer from high non-linearity as well as inconsistent cycle-to-cycle weight variation without the assistance of external circuitry, which in the case of FeFETs can results in a verification period that is microseconds in length [15]. Additionally, the physical robustness of SOT switching MTJs introduces a considerably larger endurance than 2-terminal devices, The ability to tune the change in resistance through the device geometry also provides unique ways to adapt MTJs for the circuit.

Furthermore, at the system level, the DW-MTJ-informed ACAM demonstrates the ability to perform a "fusion" of nonlinear activation and ADC. The search operation of ACAM can be used to binarize an analog input signal, while also introducing an *in-situ* nonlinearity characteristic. Thus, this eliminates the need for costly A/D converters for non-linear activation in analog computing applications. The approximation of the ReLU-alpha activation function using this concept is verified in our work, Ref. 14. There, the cost of the MACAM search operation is 0.92 fJ with an associated 0.52 fJ/search cost from the peripheral circuitry, as compared to ADC configuration of ~10.1 pJ [adc-1] and ~18.6 pJ [adc-2] in Ref. [14].

## VIII. Conclusions

Our 10-transistor, 2-DW-MTJ circuit utilizes the programmable behavior of shape-depended multi-weight DW-MTJ synapses to perform analog CAM operation. We examined the many trade-offs and design considerations in magnetic stack characteristics and device geometry in the process of designing DW-MTJs to optimize performance in ACAMs. With this, we were able to demonstrate 5 discrete multi-bit levels with realistic magnetic stack parameters, and up to 16 discrete levels using ideal projected stack parameters. The analog programmability made available by the introduction of DW-MTJs eliminates the need to interface with ADCs, which are heavily energy intensive. Additionally, the digital output enables the ACAM to also act as an alternative to ADCs. The programmable weights in ACAM makes for ideal implementation in high-throughput computing, such as one-shot/few-shot learning using decision trees.

We did not account for ACAM's intended use in large arrays to be able to handle input word lengths. This type of system-level application of ACAM is associated with changes to both average latency per search per cell, as well as energy consumption per cell. These are important considerations to be addressed in future work.